\documentclass[aps,prc,floatfix,amsmath]{revtex4-1}
\usepackage{dcolumn}
\usepackage{amssymb}
\usepackage{mathrsfs}
\usepackage{supertabular}
\usepackage{color,graphicx}
\usepackage[bookmarksopen]{hyperref}
\def  \p    {\pi}

\def  \m    {\mu}

\def  \f    {\frac}

\def  \th   {\theta}

\def  \ra   {\rightarrow}

\def  \veps {\varepsilon}

\def  \del  {\partial}

\def  \bef  {\begin{figure}}
\def  \eef  {\end{figure}}
\def  \be   {\begin{equation}}
\def  \ee   {\end{equation}}
\def  \ba   {\begin{array}}
\def  \ea   {\end{array}}
\def  \bea  {\begin{eqnarray}}
\def  \eea  {\end{eqnarray}}
\def  \beq  {\begin{eqnarray}}
\def  \eeq  {\end{eqnarray}}
\def  \nn   {\nonumber}
\def  \bd   {\begin{displaymath}}
\def  \ed   {\end{displaymath}}
\def  \bse  {\begin{subequations}}
\def  \ese  {\end{subequations}}
\def  \bwt  {\begin{widetext}}
\def  \ewt  {\end{widetext}}

\def  \ba   {{\bf{a_1}}}

\begin{document}
\title{ Next to leading order non Fermi liquid corrections to the neutrino emissivity and cooling of
the neutron star}
\author{Souvik Priyam Adhya}
\email{souvikpriyam.adhya@saha.ac.in}

\author{Pradip K. Roy}
\email{pradipk.roy@saha.ac.in}

\author{Abhee K. Dutt-Mazumder}
\email{abhee.dm@saha.ac.in}

\affiliation{High Energy Nuclear and Particle Physics Division, Saha Institute of Nuclear Physics,
1/AF Bidhannagar, Kolkata-700 064, INDIA}

\medskip

\begin{abstract}
In this work we derive the expressions of the neutrino mean free path(MFP) and emissivity with non Fermi liquid corrections up to next to leading order(NLO) in
degenerate quark matter. The calculation has been performed both for the absorption and scattering processes. 
Subsequently the role of these NLO corrections on the cooling of the neutron star has been demonstrated. 
The cooling curve shows moderate enhancement compared to the leading order(LO) non-Fermi liquid result. Although the overall correction to the MFP and emissivity are larger compared to the free Fermi gas, the cooling behavior does not alter significantly. 
\end{abstract}


\maketitle

\section{Introduction}
It is known for quite sometime now that the degenerate Fermi gas at low or zero temperature gives rise to phenomenon which is different from
the normal Fermi liquid behaviour once the magnetic interactions are included. This interesting feature is characterized by the appearance of anomalous logarithmic term in the expressions for various physical quantities like specific heat, entropy etc\cite{holstein73}. It has also been revealed recently that at strictly zero temperature, there is a logarithmic singularity in the inverse group velocity, which leads to the breakdown of the usual Fermi liquid picture in presence
of the magnetic interactions. Historically, such a deviation from the normal Fermi liquid behavior was for the first time, exposed in \cite{holstein73}, where, the specific heat of a degenerate gas due to the current-current interactions was calculated and the result contained the $T {\rm ln} T^{-1}$ term which emanates from the unscreened magnetic interactions. It is to be mentioned here that non-Fermi liquid behaviour of highly dense color superconducting QCD plasma has been studied in great detail \cite{wang02,brown00}. 

For non-relativistic systems, the magnetic interaction is suppressed in powers of $(v/c)^2$ and therefore might not be of much quantitative importance. However,
for dense plasma where the constituents like quarks or electrons are moving with a velocity close to the velocity of light, the magnetic interactions  cannot be neglected. 
In fact, it has been revealed recently that in many context, the transverse interactions, due to its infrared sensitivity, may become more important than its longitudinal
counterpart in this kinematic regime. For example, while calculating the fermion damping rate and energy loss, it has been shown in ref.\cite{Bellac97,manuel00} that the leading order (LO) contributions come from the
magnetic interaction while the longitudinal interactions contribute only at the sub leading order. In fact, it has been seen that the first two leading order
contributions in the expressions for the fermion damping rate in ultradegenerate plasma come from the transverse sector alone. Similar behaviour has also been
reported in \cite{sarkar10,sarkar11} where the authors have studied the non-Fermi liquid behaviour(NFL) of the drag and diffusion coefficient in degenerate plasma. A more elaborate discussion on the NFL aspects of the cold and dense QED and QCD plasma has been presented in \cite{boyanov01}. 

This recently discovered phenomenon of non-Fermi liquid behaviour, which relates itself to the modified quark dispersion relation for excitations close to the Fermi surface, also finds important application in astrophysics. For example, it has been shown that the NFL corrections to the quark self-energy enhance the neutrino emissivity of ungapped quark matter which may exist in the core of neutron stars\cite{schafer04,itoh70}. Like emissivity, in dense quark matter, the neutrino mean free path(MFP) also receives significant NFL corrections as has been demonstrated in
\cite{pal11}. It might be mentioned here that in all these calculations the evaluation of the quark self-energy was restricted to the leading logarithmic order. In 
\cite{rebhan05}, on the other hand, the authors determine the quark dispersion relations in ultradegenerate relativistic plasmas beyond LO, which, at zero temperature,
is characterized by the appearance of the fractional higher powers in the energy variable. In another work, the specific heat of normal degenerate quark matter has
also been calculated where also in the higher order terms the fractional powers show up \cite{ipp04}. 

In view of these contemporary investigations, we here plan to evaluate the
neutrino MFP and corresponding emissivity in normal degenerate quark matter beyond leading logarithmic order and compare with the LO results. Here leading order (LO) refers to the anomalous logarithmic term  $T {\rm log}(1/T)$ that occurs as the first term in the non-Fermi liquid contribution to the fermion self energy. Quantities such as mean free path and emissivity calculated with this term is called the LO corrections.  Next to leading order (NLO) terms include all  other terms beyond the LO that contain the fractional powers of T and up to the $(T^3){\rm log}(1/T)$ that occur in the expression of the fermion self energy. Similarly, quantities calculated with this correction are labeled as NLO corrections \cite{rebhan05}. Equipped with these results and knowing the specific heat of dense quark
matter upto the order concerned, we investigate the cooling behaviour of the neutron star with dense quark core.

The plan of the paper is as follows. In section II, we describe the formalism; where we start with the quark dispersion relations and the modifications due to NFL effects followed by evaluation of the MFP for the degenerate and non-degenerate neutrinos. These are followed by the calculation of emissivity of the neutrinos.  Section III is devoted to the study of the cooling process via neutrino emission. Finally, the results are summarized in section IV followed by conclusion in section V.
\section{Formalism}
\subsection{Quark dispersion relation}
\begin{figure}[htb]
\begin{center}
\resizebox{8.5cm}{4.75cm}{\includegraphics{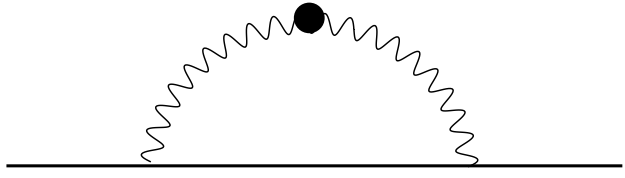}}
\caption{Fermion self-energy with resummed gluon propagator.
\label{fig1}}
\end{center}
\end{figure}
To calculate the quark dispersion relation in degenerate plasma one needs to evaluate the quark self-energy.
For this, we consider Fig.(\ref{fig1}), where the solid line represents the fermion propagator and the blob implies that the gluon propagator used here is hard dense loop (HDL) corrected propagator \cite{manuel96}. Mathematically, the quark self energy can be written as \cite{rebhan05,manuel00,brown00,ipp04,tatsumi09},
\begin{eqnarray}
\Sigma(P)= -g^2C_FT\sum_s\int{{\rm d}^3q\over (2\pi)^3}\gamma_\mu \,
S_f(i(\omega_n-\omega_s),{\bf p-q})\gamma_\nu \,
\Delta_{\mu\nu}(i\omega_s,{\bf
q}) \ ,
\label{sigma}
\end{eqnarray}
where, $P^\mu=(p^0, \bf{p})$, $p_0=i\omega_n+\mu$, $q_0=i\omega_s$, $\omega_n=(2 n+1)\pi T$ and 
$\omega_s=2\pi s T$ are the Matsubara frequencies for fermion and boson 
respectively with integers $n$ and $s$. After performing the sum over Matsubara frequency in Eq.(\ref{sigma}), $i\omega_n+\mu$ is analytically continued 
to the Minkowski space so that $P^{\mu}=(E, \bf{p})$\cite{rebhan05}. $\Sigma(P)$ can be written as a combination of quasiparticle and antiquasiparticle self energies as\cite{rebhan05},
\begin{eqnarray}
\Sigma(P)= \gamma_{0}\Lambda_{\bf{p}}^{+}\Sigma_{+}(P) - \gamma_{0}\Lambda_{\bf{p}}^{-}\Sigma_{-}(P)\nonumber\\
\end{eqnarray}
where 
\begin{eqnarray}
 \gamma_{0}S^{-1}=S_{+}^{-1}\Lambda_{\bf{p}}^{+} + S_{-}^{-1}\Lambda_{\bf{p}}^{-}\nonumber\\
S_{\pm}^{-1}=-[p^{0}\mp(|\bf{p}|+\Sigma_{\pm})]\nonumber\\
\end{eqnarray}
with the energy projection operators are given by,
\bea
\Lambda_{\bf{p}}^{\pm}=\frac{1}{2}(1\pm \gamma_{0}\gamma^{i}\hat{p}^{i})
\eea

Interactions within the medium severely modify the
on-shell self-energy of the
quarks which is manifested in the slope of the dispersion relation for the
relativistic degenerate plasma. For quasi-particles with momenta close to the
Fermi 
momentum $p_{f}(i)$ where $i$ denotes the quark flavor, the one-loop self-energy is dominated by the soft gluon 
exchanges \cite{sarkar10,manuel00}.
For the calculation of the MFP and emissivity one needs to know the modified dispersion relation which is determined as \cite{pisarski90,manuel96,tatsumi09},
\beq
\omega_{\pm}=\pm(E_{p(\omega_{\pm})} +{\rm Re}\Sigma_{\pm}(\omega_{\pm},p(\omega_{\pm})))
\eeq
where $\omega$ is the quasiparticle/antiquasiparticle energy which is a solution of the dispersion relation and $E_{p(\omega)}=\sqrt{p^{2}(\omega)+m_{q}^2}$ is the kinetic energy. As we are considering only quasiparticles, we will consider only $\omega_+$ and denote $\omega_{+}$ by $\omega$.
The above expression will be used to obtain $dp/d\omega$ needed for the phase space evaluation of the mean free path.
The authors of \cite{rebhan05} have already calculated the fermion self energy with terms beyond LO. 
We here quote the low temperature expansion of the on-shell fermion self energy for $|{\bf p}|= E$ (ultrarelativistic case) \cite{rebhan05,manuel00} and notice that no explicit dependence on the spatial momentum $\bf{p}$ occurs \cite{tatsumi09}:
\begin{eqnarray}
\Sigma_{+}(\omega)&=&{M_\infty^2\over 2E}-g^2C_Fm\,
 \Big\{{\varepsilon\over12\pi^2m}\Big[\log\Big({4\sqrt{2}m\over\pi \varepsilon}\Big)+1\Big]+{i \varepsilon\over24\pi m}
  \,+{2^{1/3}\sqrt{3}\over45\pi^{7/3}}\left({\varepsilon\over m}\right)^{5/3}(\mathrm{sgn}(\varepsilon)-\sqrt{3}i)\qquad\nn\\
  &&+ {i\over64 \sqrt{2}}\left({\varepsilon\over m}\right)^2
  -20{2^{2/3}\sqrt{3}\over189\pi^{11/3}}\left({\varepsilon\over m}\right)^{7/3}(\mathrm{sgn}(\varepsilon)+\sqrt{3}i)\qquad\nn\\
&&-{6144-256\pi^2+36\pi^4-9\pi^6\over864\pi^6}\Big({\varepsilon\over m} 
  \Big)^3 \Big[\log\left({{0.928}\,m\over \varepsilon}\right) 
-{i\pi\mathrm{sgn}(\varepsilon)\over 2}  \Big]
  +\mathcal{O}\Big(\left({\varepsilon\over m}\right)^{11/3}\Big) \Big\},
\end{eqnarray}
where $\varepsilon = (\omega-\mu)\sim T$ where NFL effects dominate. The scale of the last logarithm was determined by resumming infrared enhanced contributions.
The first term in the above expression gives the hard part contribution to the self energy where $M_\infty^{2}=g^{2}C_{F}\mu^{2}/(4\pi^{2})$ and $m$ is given by $m^{2}=N_{f}g^{2}\mu^{2}/(4\pi^{2})$ and is related to the Debye mass by $m^{2}=m^{2}_{D}/2$.
It is interesting to note here that at higher order, fractional powers in $\varepsilon$ appear. This can be attributed to the dynamical screening for the
transverse exchange of gauge bosons.

\subsection{Mean free path (MFP) of the degenerate neutrinos}
The degenerate neutrinos refers to the case where the neutrino chemical potential ($\mu_{\nu}$) is much larger than the temperature.
In the interior of a neutron star, there are two distinct phenomena for which 
the neutrino mean free path is calculated, one is absorption and the other involves scattering of neutrinos \cite{iwamoto82}. To calculate the MFP for the absorption process we consider the simplest $\beta$ decay reactions; {\em i.e} the absorption 
process and its inverse \cite{iwamoto82,iwamoto80,lattim91},
\beq
d+\nu_{e}\rightarrow u+e^-
\label{dir}
\eeq
\beq
u+e^-\rightarrow d+\nu_{e}.
\label{inv}
\eeq
 The corresponding mean free paths
are denoted by $l^{abs}_{mean}$ and $l^{scatt}_{mean}$.

The neutrino MFP is related to the total interaction rate due 
to neutrino emission averaged over the initial quark spins and summed over 
the final state phase space and spins. It is given by\cite{iwamoto82},
\bea
\label{mfp01}
\frac{1}{l_{mean}^{abs}(E_{\nu},T)}=&&\frac{g^{\prime}}{2E_{\nu}}\int\frac{d^3p_d}{
(2\p)^3}
\frac{1}{2E_d}\int\frac{d^3p_u}{(2\p)^3}
\frac{1}{2E_u}\int\frac{d^3p_e}{
(2\p)^3 }
\frac{1}{2E_e}
(2\pi)^4\delta^4(P_d +P_{\nu}-P_u
-P_e)\nn\\
&&\times|M|^2 \{n(p_d)[1-n(p_u)][1-n(p_e)]
+n(p_u)n(p_e)[1-n(p_d)]\},
\eea
where, $g^{\prime}$ is the spin and color degeneracy, 
taken to be $6$ and $E$, $p$ and $n_{p}$ are the energy, momentum and distribution function for the corresponding particle.
$|M|^2$ is the squared invariant amplitude
and is given by
$|M|^2=64G^2\cos^2\th_c(P_d\cdot P_\nu)(P_u\cdot P_e)$. Here $G\simeq1.435\times10^{-49} erg-cm^{3}$ is the weak coupling constant.
Here, we work with the two flavor system as the interaction involving strange 
quark is Cabibbo suppressed. 
We now consider the case of degenerate neutrinos 
{\em i.e.} when $\mu_{\nu}\gg T$. So in
this case both the direct Eq.(\ref{dir}) and inverse 
Eq.(\ref{inv})
processes can occur.
Consequently, the $\beta$ equilibrium condition becomes $\mu_d+\mu_{\nu}=\mu_u+\mu_e$. 
Now, to carry out the momentum integration, $d^{3}p_d$ and $d^{3}p_u$ can be evaluated as,
\bea
d^{3}p_d=2\pi \f{p_{f}(d)}{p_{f}(\nu)}pdp\f{dp_{d}}{d\omega}d\omega;
d^{3}p_u=2\pi \f{p_{f}(u)p_{f}(e)}{p}dE_{e}\f{dp_{u}}{d\omega}d\omega
\eea
where we define $p\equiv|p_{d}+p_{\nu}|=|p_{u}+p_{e}|$. $dp(\omega)/d\omega$ can be evaluated from the modified dispersion relation as follows \cite{tatsumi09},
\bea
\f{d\omega}{dp(\omega)}\simeq\f{dE_{p(\omega)}}{dp(\omega)} + \f{\del Re\Sigma_{+}(\omega)}{\del \omega}\f{d\omega}{dp(\omega)},\nonumber\\
\f{dp(\omega)}{d\omega}= \Big(1-\f{\del Re\Sigma_{+}(\omega)}{\del \omega}\Big)\f{E_{p(\omega)}}{p(\omega)}
\eea
where $\del Re\Sigma_{+}/\del p\simeq 0$, since $p$ does not appear explicitly in the expression for $\Sigma_{+}(\omega)$.
Neglecting the quark-quark interactions, the leading order result is obtained as,
\bea
\frac{1}{l_{mean}^{abs,D}}\Big|_{LO}\simeq\frac{2}{3\pi^{5}}G_{F}^{2}C_{F}\cos^{2}\th_{c}\frac{\mu_{e}^{3}}{\mu_{\nu}^{2}}\Big[1+\frac{1}{2}\Big(\frac{\mu_{e}}{\mu}
\Big)
+\frac{1}{10}\Big(\frac{\mu_{e}}{\mu}\Big)^{2}\Big]
[(E_{\nu}-\mu_{\nu})^{2}+\pi^{2} T^{2}](g\mu)^{2}\text{log}\Big(\frac{4g\mu}{\pi^{2}T}\Big).
\eea
The NLO result is evaluated as,
\bea
\frac{1}{l_{mean}^{abs,D}}\Big|_{NLO}&\simeq&\frac{8}{\pi^{3}}G_{F}^{2}C_{F}\cos^{2}\th_{c}\frac{\mu_{e}^{3}}{\mu_{\nu}^{2}}\Big[1+\frac{1}{2}\Big(\frac{\mu_{e}}{\mu}
\Big)
+\frac{1}{10}\Big(\frac{\mu_{e}}{\mu}\Big)^{2}\Big]
[(E_{\nu}-\mu_{\nu})^{2}+\pi^{2} T^{2}]\Big[a_{1}T^{2/3}(g\mu)^{4/3}\nonumber\\
&+&a_{2}T^{4/3}(g\mu)^{2/3}+a_{3}\Big\{1-3 \text{log}\Big(\frac{0.209g\mu}{T}\Big)\Big\}T^{2}\Big]
\eea
where the constants are,
\bea
a_1=\frac{2^{2/3}}{9\sqrt{3}\pi^{5/3}};
a_2=-\frac{140\times2^{4/3}}{189\sqrt{3}\pi^{7/3}}
\eea
and
\bea
a_3=\frac{6144-256\pi^{2}+36\pi^{4}-9\pi^{6}}{432\pi^{4}}.
\eea
To arrive at the Fermi-liquid result, one can use the free dispersion relation to arrive at,
\bea
\label{mfp_cond1}
\frac{1}{l_{mean}^{abs,D}}\Big|_{FL}=\frac{4}{\pi^{3}}G_{F}^{2}\cos^{2}\th_{c}
\frac{\mu^{2}
\mu_{e}^{3}}{\mu_{\nu}^{2}}\Big[1+\frac{1}{2}\Big(\frac{\mu_{e}}{\mu}
\Big)
+\frac{1}{10}\Big(\frac{\mu_{e}}{\mu}\Big)^{2}\Big]
[(E_{\nu}-\mu_{\nu})^{2}+\pi^{2} T^{2}].
\eea 
 Since quarks and electrons are assumed to be massless,
the chemical equilibrium condition gives $p_f(u)+p_f(e)=p_f(d)+p_f(\nu)$,
which we use to derive Eq.(\ref{mfp_cond1}). We have further assumed that $\mu_d \sim \mu_u =\mu$.
Next we calculate the MFP for the quark-neutrino scattering process,
\beq
q_{i}+\nu_e({\overline \nu_e})\ra q_{i}+\nu_e({\overline \nu_e})
\eeq
for each quark component of flavor $i (=u~{\rm or}~d)$.
Including the NFL corrections through the phase space and assuming $m_{q_i}/p_{f_i}\ll 1$, we obtain,
\bea
\frac{1}{l_{mean}^{scatt,D}}\Big|_{FL}=\f{3}{4\pi}n_{q_i}G_{F}^{2}
\times[(E_{\nu}
-\mu_ { \nu } )^ { 2 } +\pi^{2} T^{2}]\Lambda(x_i);
\eea
\bea
\frac{1}{l_{mean}^{scatt,D}}\Big|_{LO}\simeq\f{1}{8\pi^{3}}n_{q_i}C_{F}G_{F}^{2}[(E_{\nu}
-\mu_ { \nu } )^ { 2 } +\pi^{2} T^{2}]\Lambda(x_i)g^{2}\text{log}\Big(\f{4g\mu}{\pi^{2}T}\Big);
\eea
\bea
\frac{1}{l_{mean}^{scatt,D}}\Big|_{NLO}&\simeq&\f{3}{2\pi}n_{q_i}C_{F}G_{F}^{2}[(E_{\nu}
-\mu_ { \nu } )^ { 2 } +\pi^{2} T^{2}]\Lambda(x_i)\Big[a_{1}g^{4/3}\Big(\f{T}{\mu}\Big)^{2/3}\nonumber\\
&+&a_{2}g^{2/3}\Big(\f{T}{\mu}\Big)^{4/3}+a_{3}\Big\{1-3\text{log}\Big(\frac{0.209g\mu}{T}\Big)\Big\}\Big(\f{T}{\mu}\Big)^{2}\Big]
\eea
where $n_{q_{i}}$ is the number density of quark of flavor i, given by,
\beq
n_{q_i}&=&6\int\frac{d^3p}{(2\p)^3}\frac{1}{e^{\beta(E_{q_i}-\mu_{q_i})}}
\eeq
where $6$ is the quark degeneracy factor.
$m_{q_{i}}$ is the mass of
quark, $\sigma_0 \equiv 4G_{F}^2 m_{e}^2/\pi$ \cite{tubb75,lamb76}
 and $\Lambda(x_i)$ is defined in
\cite{pal11} where $x_i=\mu_{\nu}/\mu_{q_i}$ if $\mu_{\nu}< \mu_{q_i}$
and $x_i=\m_{q_i}/\mu_{\nu}$ if $\mu_{\nu}> \mu_{q_i}$.
The contributions from the Fermi liquid(FL), LO and NLO are added to obtain the MFP for the corresponding process.
Further, one can combine $l^{scatt}_{mean}$ with $l^{abs}_{mean}$ 
to define
total mean free path as \cite{sagert06},
\beq
\frac{1}{l_{mean}^{total}}&=&\frac{1}{l_{mean}^{abs}}+\frac{1}{l_{mean}^{scatt}}.
\eeq 
In addition to the inclusion of the NLO terms, the Fermi liquid term and LO term agrees with that of \cite{pal11}.
\subsection{Mean free path of non-degenerate neutrinos}
We now derive MFP for nondegenerate neutrinos 
{\em i.e.} when $\mu_{\nu}\ll T$ beyond the Fermi-liquid contribution. For nondegenerate 
neutrinos the inverse process (\ref{inv}) is dropped. We are considering only depopulation of neutrinos as in non-degenerate (untrapped) case repopulation or the reverse reaction is assumed to be zero. Hence, we neglect
the second term in the curly braces of Eq.(\ref{mfp01})\cite{iwamoto80,iwamoto82}. For free quarks, 
the matrix element vanishes \cite{iwamoto82,duncan83,huang07}, since $u$, $d$ quarks and electrons are collinear in
momenta. The inclusion of strong interactions between quarks relaxes
these kinematic restrictions resulting in a nonvanishing squared matrix amplitude. We can neglect the neutrino momentum in energy conserving relation due to the thermal production of the neutrinos \cite{iwamoto82}. Following the procedure described in \cite{tatsumi09,shapiro_book,haensel01} the MFP
for the Fermi liquid case, LO and NLO are obtained as,
\bea\label{mfp03}
\frac{1}{l_{mean}^{abs,ND}}\Big|_{FL}&=&\frac{3C_F\alpha_s}{\pi^4}G_{F}^2\cos^2\th_c
~\mu_d~\mu_u~\mu_e~\frac{(E_{\nu}^2+\pi^2 T^2)}{(1+e^{-\beta E_{\nu}})};
\eea
\bea
\frac{1}{l_{mean}^{abs,ND}}\Big|_{LO}\simeq\frac{C_{F}^{2}\alpha_s}{2\pi^6}G_{F}^2\cos^2\th_c\mu_{e}\frac{(E_{\nu}^2+\pi^2 T^2)}{(1+e^{-\beta E_{\nu}})}(g\mu)^{2}\text{log}\Big(\f{4g\mu}{\pi^{2}T}\Big);
\eea
\bea
\frac{1}{l_{mean}^{abs,ND}}\Big|_{NLO}&\simeq&\frac{3C_{F}^{2}\alpha_s}{\pi^4}G_{F}^2\cos^2\th_c\mu^{2}\mu_{e}\frac{(E_{\nu}^2+\pi^2 T^2)}{(1+e^{-\beta E_{\nu}})}\Big[b_{1}g^{4/3}\Big(\f{T}{\mu}\Big)^{2/3}\nonumber\\
&+&b_{2}g^{2/3}\Big(\f{T}{\mu}\Big)^{4/3}+b_{3}\Big\{1-3\text{log}\Big(\f{0.209g\mu}{T}\Big)\Big\}\Big(\f{T}{\mu}\Big)^{2}\Big]
\eea
where the constants are evaluated as,
\bea
b_{1}=\f{2^{5/3}}{9\sqrt{3}\pi^{5/3}};
b_2=-\f{280\times2^{4/3}}{189\sqrt{3}\pi^{7/3}}
\eea
and
\bea
b_3=\frac{6144-256\pi^{2}+36\pi^{4}-9\pi^{6}}{216\pi^{4}}.
\eea
Similarly, for the scattering of nondegenerate neutrinos in quark matter with appropriate phase space corrections we obtain,
\bea\label{mfp_scnd}
\frac{1}{l_{mean}^{scatt,ND}}\Big|_{FL}&=&\frac{C_{V_{i}}^2 +
C_{A_i}^2}{5\pi}n_{q_i}G_{F}^{2}\f{E_{\nu}^{3}}{\mu};
\eea
\bea
\frac{1}{l_{mean}^{scatt,ND}}\Big|_{LO}&\simeq&\frac{C_{V_{i}}^2 +
C_{A_i}^2}{30\pi^{3}}n_{q_i}G_{F}^{2}C_{F}\f{E_{\nu}^{3}}{\mu}g^{2}\text{log}\Big(\f{4g\mu}{\pi^{2}T}\Big);
\eea
\bea
\frac{1}{l_{mean}^{scatt,ND}}\Big|_{NLO}&\simeq&(C_{V_{i}}^2 + C_{A_i}^2)n_{q_i}G_{F}^{2}C_{F}\Big[b_{1}'\f{T^{2/3}g^{4/3}}{\mu^{5/3}}+b_{2}'\f{T^{4/3}g^{2/3}}{\mu^{7/3}}+b_{3}'\Big\{1-3\text{log}\Big(\f{0.209g\mu}{T}\Big)\Big\}\Big(\f{T^2}{\mu^3}\Big)\Big],
\eea
where the constants are,
\bea
b_{1}'=\f{2^{5/3}}{45\sqrt{3}\pi^{8/3}};
b_{2}'=-\f{56\times2^{4/3}}{189\sqrt{3}\pi^{10/3}}
\eea
and
\bea
b_{3}'=\frac{6144-256\pi^{2}+36\pi^{4}-9\pi^{6}}{1080\pi^{5}}.
\eea
Here, we have assumed $m_{q_i}/p_{f_i}\ll 1$.
Thus, the total MFP for non-degenerate neutrinos is obtained by summing up the contributions from the absorption and scattering parts to get the expression of the MFP of the non-degenerate neutrinos up to the NLO terms.
So, we get the expression of the MFP of the non-degenerate neutrinos up to the NLO terms.
\subsection{Emissivity of non-degenerate neutrinos}
The total emissivity of the non-degenerate neutrinos is obtained by multiplying the neutrino energy with the inverse of the MFP with appropriate factors and integrated over the neutrino momentum.
The relation between neutrino emissivity and the neutrino mean free path is thus obtained as \cite{sawyer},
\bea
\veps=\int \frac{d^{3}p_{\nu}}{(2\pi)^{3}}E_{\nu}\frac{1}{l(-E_{\nu},T)}.
\eea
Using the mean free path for the non-degenerate neutrinos we obtain,
\bea
\varepsilon - \varepsilon_{0} = \varepsilon_{LO} + \varepsilon_{NLO}
\eea
where,
\bea
\varepsilon_{0} \simeq \frac{457}{630}G_{F}^{2}cos^{2}\theta_{c}\alpha_{s}\mu_{e}T^{6}\mu^2
\eea
is the usual Fermi liquid contribution which agrees with the result presented in ref.\cite{iwamoto82}.
At the LO we have obtained,
\bea
\varepsilon_{LO} \simeq \frac{457}{3780}G_{F}^{2}cos^{2}\theta_{c}C_{F}\alpha_{s}\mu_{e}T^{6}\frac{(g\mu)^2}{\pi^2}\text{ln}\Big(\frac{4g\mu}{\pi^{2}T}\Big)
\eea
which is in agreement with the result quoted in ref.\cite{schafer04}.
Now, following the procedure in \cite{schafer04,pal11}, we obtain the NLO contribution to the neutrino emissivity as,
\bea
\varepsilon_{NLO} \simeq \frac{457}{315}G_{F}^{2}cos^{2}\theta_{c}C_{F}\alpha_{s}\mu_{e}T^{6}\Big[c_{1}T^{2} + c_{2}T^{2/3}(g\mu)^{4/3} - c_{3}T^{4/3}(g\mu)^{2/3} - c_{4}T^{2}\text{ln}\Big(\frac{0.656g\mu}{\pi T}\Big)\Big]
\eea
where the constants are evaluated as,
\bea
c_1 = -0.0036\pi^{2};
c_2 = \frac{2^{2/3}}{9\sqrt{3}\pi^{5/3}};
c_3 = \frac{40\times2^{1/3}}{27\sqrt{3}\pi^{7/3}}
\eea
and
\bea
c_4 = \frac{6144-256\pi^{2}+36\pi^{4}-9\pi^{6}}{144\pi^{4}}.
\eea
The NFL correction only appear in the phase space integral of the MFP \cite{pal11} and subsequently in the expression of the emissivity \cite{schafer04}. It is actually related to the unscreened transverse interaction \cite{manuel00}.
The factor of $T^6$ can be understood easily. Naively, we can see that one power of $T$ is obtained from phase space integral of a degenerate fermion. Further, we obtain a $T^3$ from the phase space integral of the neutrino. One power of $T$ from the energy conserving $\delta$ function is cancelled by a power from the emitted neutrino energy. It is to be noted that for excitations near the Fermi surface, the angular integrals give no temperature dependence.
Now putting the standard values 
\cite{schafer04,pal11} for the parameters, the corrections can be compared with the results given in \cite{schafer04}. 
\section{Cooling process via neutrino emission}
\begin{figure}[]
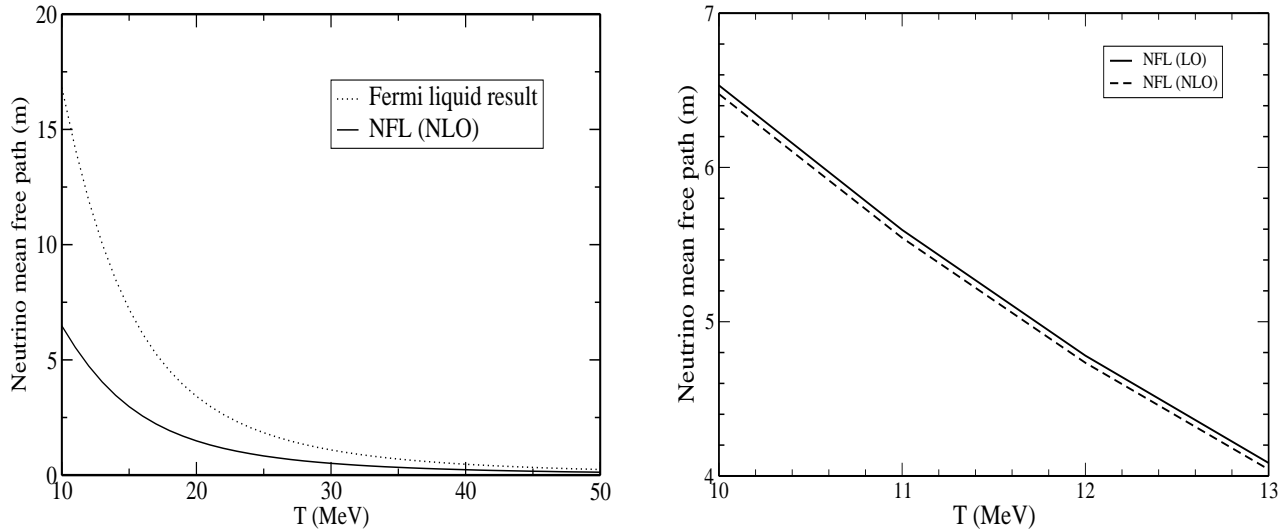

\bigskip
\begin{center}
\resizebox{8.0cm}{7.0cm}{\includegraphics{mfp_d.eps}}
~~~~~~~\resizebox{8.0cm}{7.0cm}{\includegraphics{mfp_d1.eps}}
\caption{Mean free path of degenerate neutrinos.The left panel shows a comparison between the Fermi liquid result and NLO corrections for the non-Fermi liquid effects. The right panel shows the reduction of the MFP due to NLO corrections.
\label{fig2}}
\end{center}
\end{figure}
The temperature of the neutron star with a quark matter core shows a dependency with time. To analyse the cooling of the star \cite{peth04,finley92,rutledge01}, the specific heat capacity of the quark matter core needs to be taken into consideration along with the emissivity via the cooling process \cite{iwamoto82,heisel},
\beq
  \frac{\partial u}{\partial t} = \frac{\partial u}{\partial T} 
  \frac{\partial T}{\partial t} = c_v(T) \frac{\partial T}{\partial t} 
    = - \varepsilon(T),
\label{eq_cool}
\eeq
where $u$ is the internal energy, $t$ is time and we have assumed 
that there is no surface emission. The NFL effects on the specific heat capacity of the degenerate quark matter has been calculated recently in \cite{ipp04}.  
The specific heat capacity for a non-color-superconducting degenerate quark matter is given as \cite{ipp04},
\beq
\label{spec-heat}
\label{finalcv}
  &&\!\!\!\!\!\!\!\!\!\!\!\!
  {\mathcal C_v-\mathcal C_v^0\over N_g}={g_{eff}^2\mu^2 T\over36\pi^2}\left(\ln\left({4g_{eff}\mu\over\pi^2T}\right)+\gamma_E
  -{6\over\pi^2}\zeta^\prime(2)-3\right)\nonumber\\
  &&-40{2^{2/3}\Gamma\left({8\over3}\right)\zeta\left({8\over3}\right)\over27\sqrt{3}\pi^{11/3}}
  T^{5/3}(g_{eff}\mu)^{4/3}
  +560{2^{1/3}\Gamma\left({10\over3}\right)\zeta\left({10\over3}\right)
  \over81\sqrt{3}\pi^{13/3}}T^{7/3}(g_{eff}\mu)^{2/3}\nonumber\\
  &&+{2048-256\pi^2-36\pi^4+3\pi^6\over180\pi^2}T^3
  \left[\ln\left({g_{eff}\mu\over T}\right)+\bar c-{\frac{7}{12}}\right]\nonumber\\
&&+{O}(T^{11/3}/(g_{eff}\mu)^{2/3})
+{O}(g^4\mu^2 T \ln T)  \label{cv1},
\eeq
where the coupling constant $g$ is related to $g_{eff}$ as,
\begin{equation}\label{geffdef}
g^2 = \frac{2\ g^{2}_{eff}}{N_{f}},
\end{equation}
and $N_{f}$ is the number of quark flavors. The contribution from the free part is given as,
\beq
\mathcal C_{v}^{0}=N N_f \frac{\mu^{2}T}{3}. 
\eeq
Using the above expression for specific heat and the emissivity expression upto NLO, we analyze the cooling behavior of the neutron star.
\section{Results}
\begin{figure}[]
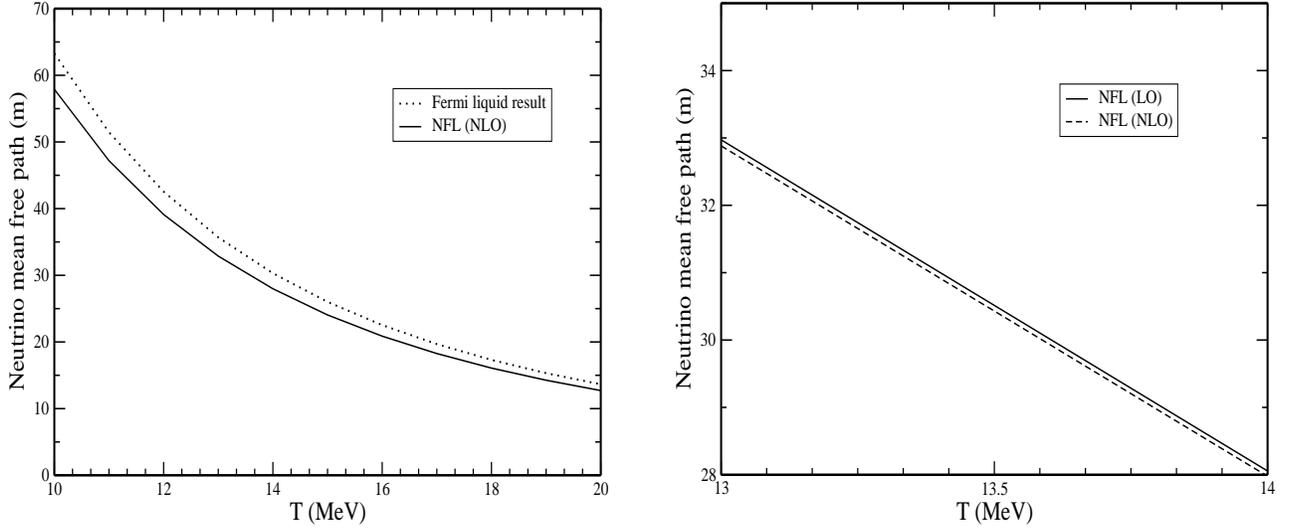

\begin{center}
\resizebox{8.0cm}{7.0cm}{\includegraphics{mfp_nd.eps}}
~~~~~~~\resizebox{8.0cm}{7.0cm}{\includegraphics{mfp_nd1.eps}}
\caption{Same as Fig.\protect\ref{fig2} for non-degenerate neutrinos.}
\label{fig3}
\end{center}
\end{figure}
\begin{figure}[htb]
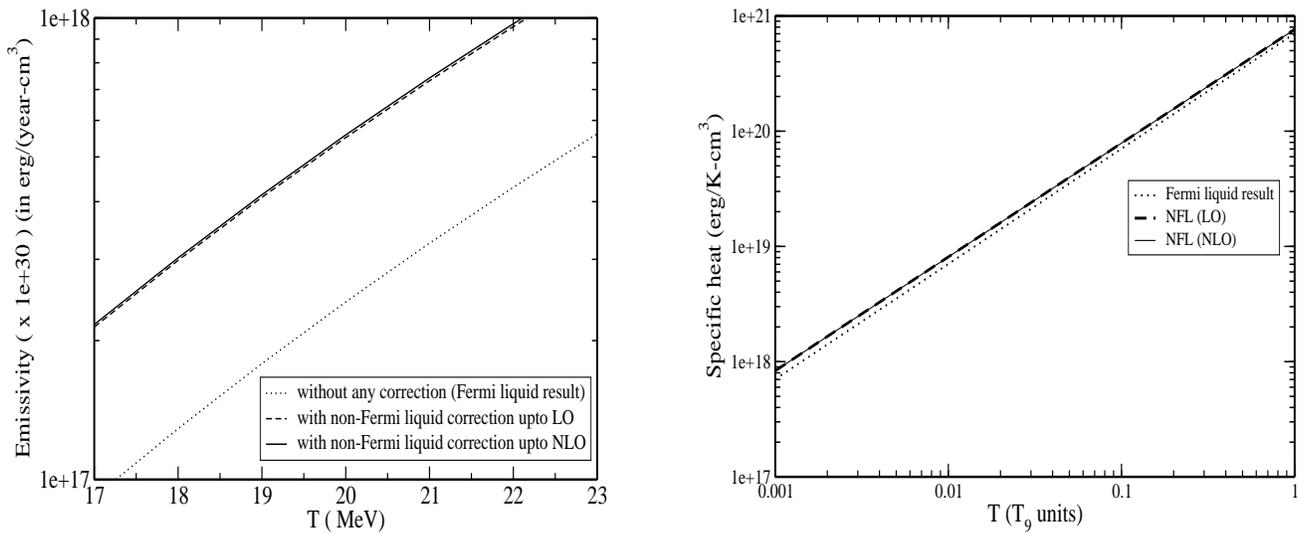

//
\begin{center}
\resizebox{8.0cm}{7.0cm}{\includegraphics{emissivityv1.eps}}
~~~~~~~~~~\resizebox{8.0cm}{7.0cm}{\includegraphics{cv_pkr.eps}}
\caption{The left panel shows the emissivity of the neutrinos with temperature in degenerate quark matter. The right panel shows the behavior of the specific heat of the degenerate quark matter with temperature ($T_9$ in units of $10^9$ K).
\label{fig4}}
\end{center}
\end{figure}
\bigskip
\begin{figure}[htb]
\begin{center}
\resizebox{10.0cm}{7.0cm}{\includegraphics{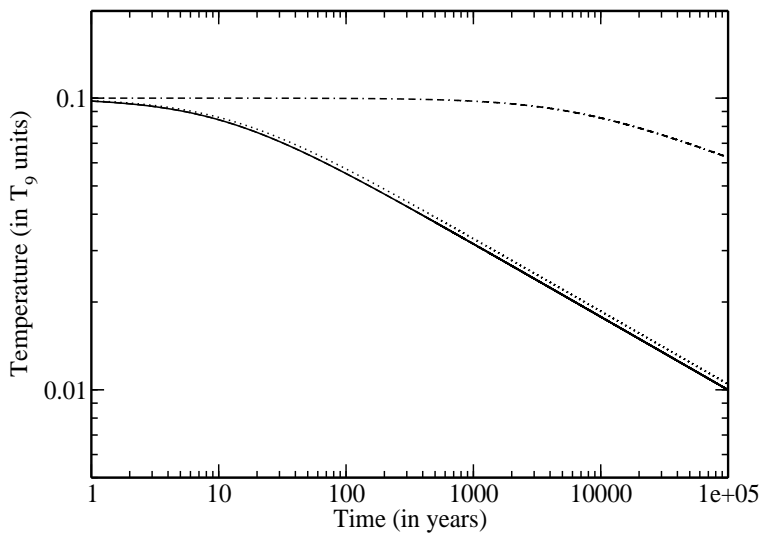}}
\caption{The cooling behavior of neutron star with core as neutron matter and degenerate quark matter. The dotted line represents the Fermi liquid result, 
the solid line represents the non-Fermi NLO correction. 
The dash-dotted line gives the cooling behavior of  
the neutron star core made up of purely neutron matter. 
\label{fig5}}
\end{center}
\end{figure}
An estimation of the MFP of neutrinos with the temperature has been presented in this section. For this purpose, we have assumed the quark chemical potential to be $500$ MeV. This is in well agreement with the high density $\sim 6\rho_{0}$ ($\rho_{0}$ being nuclear matter saturation density) at the core.
The chemical potential may well have a time dependence. This would gain significance for terms beyond order $T^{3}$ since $T/(g\mu)<<1$. 
We have also taken $\mu_{e}=15 MeV$ and $\alpha_{s}=0.1$. 
In left panel of Fig.(\ref{fig2}) we note 
there is a considerable decrease in MFP of degenerate neutrinos due to 
NLO corrections over the Fermi liquid result. 
In right panel of Fig.(\ref{fig2}) LO correction is compared with 
the NLO correction and it is seen that the MFP with NLO correction
is reduced marginally as compared to LO correction.
Similar features have been observed in case of the MFP of non-degenerate
neutrinos as displayed in Fig.(\ref{fig3}). This marginal difference
in the MFP between NFL LO and NLO corrections also leads to marginal
difference in the emissivity for the two cases. This is shown in 
Fig.(\ref{fig4}). 
These small reductions are reflected in the marginally enhanced 
emissivity of the non-degenerate neutrinos which has been shown in 
left panel of Fig.(\ref{fig4}). We find that there is a modest increase 
in the emissivity of the neutrinos. The right panel of Fig.(\ref{fig4}) 
gives a comparison of the NFL corrections to the specific heat already reported in \cite{ipp04}. 
The complicated cooling equation cannot be solved analytically and we have resorted to numerical calculation.
We observe that the cooling of neutron star is marginally faster in case of NFL (NLO) as compared to the Fermi liquid result (shown in Fig.(\ref{fig5})).   
\section{Summary and Discussions}
In this work, we have calculated the MFP of degenerate and non-degenerate 
neutrinos both for the scattering and absorption processes. 
We then find the expression for neutrino emissivity for non-degenerate 
neutrinos with NLO corrections. It is seen that both MFP and 
emissivity contain terms at the higher order which 
involve fractional powers in $(T/\mu)$.
We have found that there is a decrease in the MFP due to NLO corrections. 
We reconfirm that the leading order correction to the quantities like MFP or emissivity are significant compared to the Fermi liquid results. The NLO corrections, which we derive here, have however been found to be numerically close to the LO results. We have also examined the cooling behavior of a neutron star by incorporating NLO correction to the specific heat and emissivity which affect the results considerably compared to the simple Fermi liquid case.
\section{Acknowledgments}
One of the authors [SPA] would like to thank Sreemoyee Sarkar and Mahatsab Mandal for useful discussions regarding different aspects of the paper and would like to thank UGC, India (Sr. no.: 2120951147) for providing the fellowship during the tenure of this work.


\begin{thebibliography}{50}
\bibitem{holstein73} T. Holstein, R.E. Norton and P. Pincus, Phys. Rev. B{\bf 8}, 2649 (1973).
\bibitem{wang02} Q. Wang and D. Rischke, Phys.Rev.D {\bf 65}, 
054005 (2002).
\bibitem{brown00} W.E.Brown, J.T.Liu and H.C.Ren, Phys.Rev.D {\bf 61}, 114012
(2000); {\bf 62}, 054013 (2000).
\bibitem{Bellac97} M. Le Bellac and C. Manuel, Phys. Rev. D {\bf 55}, 3215(1997).
\bibitem{manuel00} C.Manuel, Phys.Rev.D {\bf 62}, 076009 (2000).
\bibitem{sarkar10} S.Sarkar and A.K.Dutt-mazumder, Phys.Rev.D {\bf 82}, 
056003 (2010).
\bibitem{sarkar11} S.Sarkar and A.K.Dutt-Mazumder, Phys.Rev.D {\bf 84}, 
096009 (2011).
\bibitem{boyanov01} D.Boyanovsky and H.J.de Vega, Phys.Rev.D {\bf 63}, 034016 (2001).
\bibitem{schafer04} T.Sch\"{a}fer and K.Schwenzer, Phys.Rev.D {\bf 70},  
114037 (2004).
\bibitem{itoh70} N.Itoh, Prog.Theor.Phys {\bf 44}, 291 (1970).
\bibitem{pal11} K.Pal and A.K.Dutt-Mazumder,Phys.Rev.D {\bf 84}, 034004 (2011).
\bibitem{rebhan05} A.Gerhold and A.Rebhan, Phys.Rev.D {\bf 71}, 085010 (2005).
\bibitem{ipp04} A.Gerhold, A.Ipp and A.Rebhan, Phys.Rev.D {\bf 70}, 105015 (2004); {\bf 69}, R011901(2004).
\bibitem{manuel96} C.Manuel, Phys.Rev.D {\bf 53}, 5866 (1996).
\bibitem{tatsumi09} K.Sato and T.Tatsumi, Nucl.Phys.A {\bf 826}, 74 (2009).
\bibitem{pisarski90} R.D.Pisarski, Phys.Rev. Lett. {\bf 63}, 1129 (1989);
E.Braaten and R.D.Pisarski, Nucl.Phys.B {\bf 337},569 (1990).
\bibitem{iwamoto82} N.Iwamoto, Ann.Phys.(N.Y.){\bf 141}, 1 (1982).
\bibitem{iwamoto80} N.Iwamoto, Phys.Rev.Lett. {\bf 44}, 1637 (1980).
\bibitem{lattim91} J.M.Lattimer, C.J.Pethick, M.Prakash and P.Haensel,
Phys.Rev.Lett.{\bf 66}, 2701 (1991).
\bibitem{tubb75} D.L.Tubbs and D.N.Schramm, Astrophys.J.{\bf 201}, 467 (1975).
\bibitem{lamb76} D.Q.Lamb and C.J.Pethick, Astrophys.J.Lett.{\bf 209}, L77 (1976).
\bibitem{sagert06} I.Sagert and J.Schaffner-Bielich, astro-ph/0612776.
\bibitem{duncan83} R.C.Duncan, S.L.Shapiro and I.Wasserman,
Astrophys.J.{\bf 267}, 358 (1983).
\bibitem{huang07} X.Huang, Q.Wang and P.Zhuang, Phys.Rev.D {\bf 76},
094008 (2007).
\bibitem{shapiro_book} S.L.Shapiro and S.A.Teukolsky, {\it Black Holes, 
White Dwarfs and Neutron Stars.} Wiley-Interscience, New York (1983).

\bibitem{haensel01} D.G.Yakovlev, A.D.Kaminker, O.Y.Gnedin and P.Haensel,
Phys.Rept. {\bf 354}, 1-155 (2001).
\bibitem{sawyer} R. F. Sawyer and A. Soni, Astrophys. J. {\bf 216},
73 (1977).
\bibitem{peth04} D.G.Yakovlev and C.J.pethick, Ann.Rev.Astron.Astrophys. 
{\bf 42} 169 (2004).
\bibitem{finley92} J.P.Finley et al, The astrophysical Journal {\bf 394}, 
L21 (1992).
\bibitem{rutledge01} R.E.Rutledge et al, The astrophysical Journal {\bf 551}, 
921 (2001).
\bibitem{heisel} H. Heiselberg and C. J. Pethick, Phys.Rev.D {\bf 48}, 2916 (1993).



















\end{thebibliography}
\end{document}